\documentclass[twocolumn]{aastex63}
\usepackage{xspace,amsmath,amssymb,lineno,multirow}

\shorttitle{The HEROES Catalog}
\shortauthors{Taylor et al.}

\begin{document}

\title{HEROES: The Hawaii eROSITA Ecliptic Pole Survey Catalog}

\correspondingauthor{A.~J.~Taylor}
\email{ataylor@astro.wisc.edu}

\author[0000-0003-1282-7454]{A.~J.~Taylor}
\affiliation{Department of Astronomy, University of Wisconsin--Madison,
475 N. Charter Street, Madison, WI 53706, USA}

\author[0000-0002-3306-1606]{A.~J.~Barger}
\affiliation{Department of Astronomy, University of Wisconsin--Madison,
475 N. Charter Street, Madison, WI 53706, USA}
\affiliation{Department of Physics and Astronomy, University of Hawaii,
2505 Correa Road, Honolulu, HI 96822, USA}
\affiliation{Institute for Astronomy, University of Hawaii, 2680 Woodlawn Drive,
Honolulu, HI 96822, USA}

\author[0000-0002-6319-1575]{L.~L.~Cowie}
\affiliation{Institute for Astronomy, University of Hawaii,
2680 Woodlawn Drive, Honolulu, HI 96822, USA}

\author[0000-0002-0797-0646]{G.~Hasinger}
\affiliation{European Space Astronomy Centre (ESA/ESAC),
E-28691 Villanueva de la Ca\~nada, Madrid, Spain}

\author{E.~M.~Hu}
\affiliation{Institute for Astronomy, University of Hawaii,
2680 Woodlawn Drive, Honolulu, HI 96822, USA}

\author{A.~Songaila}
\affiliation{Institute for Astronomy, University of Hawaii,
2680 Woodlawn Drive, Honolulu, HI 96822, USA}

\begin{abstract}
We present a seven band ($g$, $r$, $i$, $z$, $y$, NB816, NB921) catalog derived from a Subaru Hyper Suprime-Cam (HSC) imaging survey of the North Ecliptic Pole (NEP). The survey, known as HEROES, consists of 44~deg$^2$ of contiguous imaging reaching median 5$\sigma$ depths of $g$: 26.5, $r$: 26.2, $i$: 25.7, $z$: 25.1, $y$: 23.9, NB816: 24.4, NB921: 24.4~mag. We reduced these data with the HSC pipeline software \texttt{hscPipe}, and produced a resulting multiband catalog containing over 25 million objects. We provide the catalog in three formats: (1) a collection of \texttt{hscPipe} format forced photometry catalogs, (2) a single combined catalog containing every object in that dataset with selected useful columns, and (3) a smaller variation of the combined catalog with only essential columns for basic analysis or low memory machines. The catalog uses all the available HSC data on the NEP and may serve as the primary optical catalog for current and future NEP deep fields from instruments and observatories such as SCUBA-2, eROSITA, Spitzer, Euclid, and JWST. 

\end{abstract}
\section{Introduction} \label{sec:intro}

Since its first light in 2013, Hyper Suprime-Cam \citep[HSC;][]{miyazaki18} on the Subaru 8.2m Telescope has been the premier wide field optical imager on 6--10~meter class telescopes. The large collecting area of Subaru and the wide field of view of HSC (1\fdg5) allow for the efficient observation of both wide and deep surveys. The largest of these projects---the HSC Subaru Strategic Program (HSC-SSP)---has fully mapped 670~deg$^2$ of sky in $grizy$ broadband filters and over 1470~deg$^2$ in partially observed (incomplete coverage in $grizy$) area \citep{aihara18,aihara19,aihara22}. HSC-SSP has also produced deep observations in the DEEP2-3, SXDS+XMM-LSS, ELAIS-N1, and COSMOS fields totaling $>$30~deg$^2$ of imaging in $grizy$ broadband filters, as well as NB387, NB816, and NB921 narrowband filters, all at a 5$\sigma$ depth $>$25~mag.

In 2015, given the enormous potential of HSC and its available narrowband filters, we designed and began to execute HEROES: the Hawaii EROsita Ecliptic pole Survey. Our goal was to produce a survey of the North Ecliptic Pole (NEP)
that would match the filter coverage and imaging depths of the HSC-SSP Deep fields with a much wider contiguous area,
as well as complement the then-future deepest eROSITA X-ray observations \citep{merloni12}. 

In recent years, the NEP has become a major focus of a number of ground and space-based surveys and missions spanning sub-millimeter to X-ray wavelengths. HEROES provides complementary wide-field optical broadband and narrowband data to these surveys.

At sub-millimeter wavelengths, the NEP has also been extensively observed with SCUBA-2 on the James Clerk Maxwell Telescope through the S2CLS \citep[0.6~deg$^2$, 850~$\mu$m;][]{geach17}, NEPSC2 \citep[2~deg$^2$, 850~$\mu$m;][]{shim20}, and S2TDF \citep[0.087~deg$^2$, 850$\mu$m;][]{hyun23} surveys. HEROES will provide sub-millimeter studies with corresponding optical data that may be used to find optical counterparts for sub-millimeter bright sources. The combination of optical/NIR and sub-millimeter flux may better constrain photometric redshifts and spectral energy distribution fittings for the determination of stellar mass, star formation rate, age, and dust attenuation for such sources (e.g. S.~McKay~et~al.~submitted).

In the infrared, the NEP contains the Spitzer IRAC Dark Field \citep{krick08,krick09,frost09}, as well as the upcoming 20~deg$^2$ Euclid Deep Field North \citep{amendola13,amendola18}. HEROES will serve as a natural complement and extension of these datasets into optical wavelengths, providing broadband coverage from 0.4--0.8~$\mu$m (when combined with Spitzer/IRAC) and providing individual $griz$ magnitudes to help better complement the upcoming single 550--900~nm very-broadband Euclid/VIS observations and 0.92--2~$\mu$m Euclid/NISP $YJH$ photometric and 1.1--2~$\mu$m slitless spectroscopic observations \citep{laureijs11}. The HEROES data will permit improved SED fitting for Euclid detected sources and ultimately provide a more robust understanding of the properties of these objects.

In x-rays, the NEP is already home to the deepest eROSITA X-ray observations \citep{merloni12}, and will contain the future SPHEREx Deep North Field \citep{dore16, dore18}. For these missions, HEROES and may provide insight into optical properties of x-ray detected AGN including photometric redshifts, as well as provide superior target astrometry when compared to the x-ray measurements. For example, \cite{radzom22} used Chandra data in conjunction with HSC data in the SSA22 field to produce x-ray luminosity functions at redshifts $z=0.2--4$. With the combination of HEROES, eROSITA, SPHEREx, and Euclid, this type of study could be replicated in the NEP with $>500\times$ the survey area.

Moreover, HEROES and the NEP are especially important for space-based observatories that orbit the Sun-Earth L2 Lagrange point, as the NEP and the South Ecliptic Pole are typically part of any such observatory's continuous viewing zone. As such, the NEP contains the aforementioned Sptizer IRAC Dark Field, eROSITA Deep Field, upcoming Euclid Deep Field North, and SPHEREx Deep Field North as well as the JWST Time Domain Field \citep[TDF; part of the PEARLS project;][]{windhorst17,windhorst22}. 

HEROES was initially reduced in 2017 using the Pan-STARRS Image Processing Pipeline \citep[IPP;][]{magnier16,magnier20a,magnier20b}. This version of HEROES had incomplete coverage in the $r$ and $y$ bands at low R.A.\ pointings and did not have the additional JWST TDF pointing (see \S{\ref{sec:observations}} below). 

HEROES is the largest contiguous HSC narrowband survey to date. As such, we previously used the wide-field narrowband coverage in the initial HEROES dataset for studies of Lyman-$\alpha$ Emitters (LAEs) near the epoch of reionization at $z=5.7-6.6$ \citep{songaila18,songaila22,taylor20,taylor21}, as well as the development of a broadband selection technique for emission line galaxies \citep{rosenwasser22}. 

In 2021, we completed the final HEROES observations. Here we present the photometric catalog of the complete and newly re-reduced version of HEROES, which also incorporates all the archival HSC data on this field.

In Section~\ref{sec:observations}, we describe the HSC data. In Section~\ref{sec:reduction}, we summarize the data reduction and processing. In Section~\ref{sec:catalog}, we present the final catalog and describe its format and availability. In Section~\ref{sec:quality}, we verify the catalog's quality and measure its depth in each filter. Finally, in Section~\ref{sec:samples}, we demonstrate both a $z=6.6$ LAE sample selection and $g$, $r$, and $i$-band dropout selections. 

We assume $\Omega_M=0.3$, $\Omega_\Lambda=0.7$, and $H_0=70$~km~s$^{-1}$~Mpc$^{-1}$ throughout. All magnitudes are given in the AB magnitude system, where an AB magnitude is defined by $m_{AB}=-2.5\log f_\nu - 48.60$. Here $f_\nu$ is the flux of the source in units of ergs~cm$^{-2}$~s$^{-1}$~Hz$^{-1}$.

\section{Observations}\label{sec:observations}
\begin{figure}[ht]
\centering
\includegraphics[angle=0,width=\columnwidth]{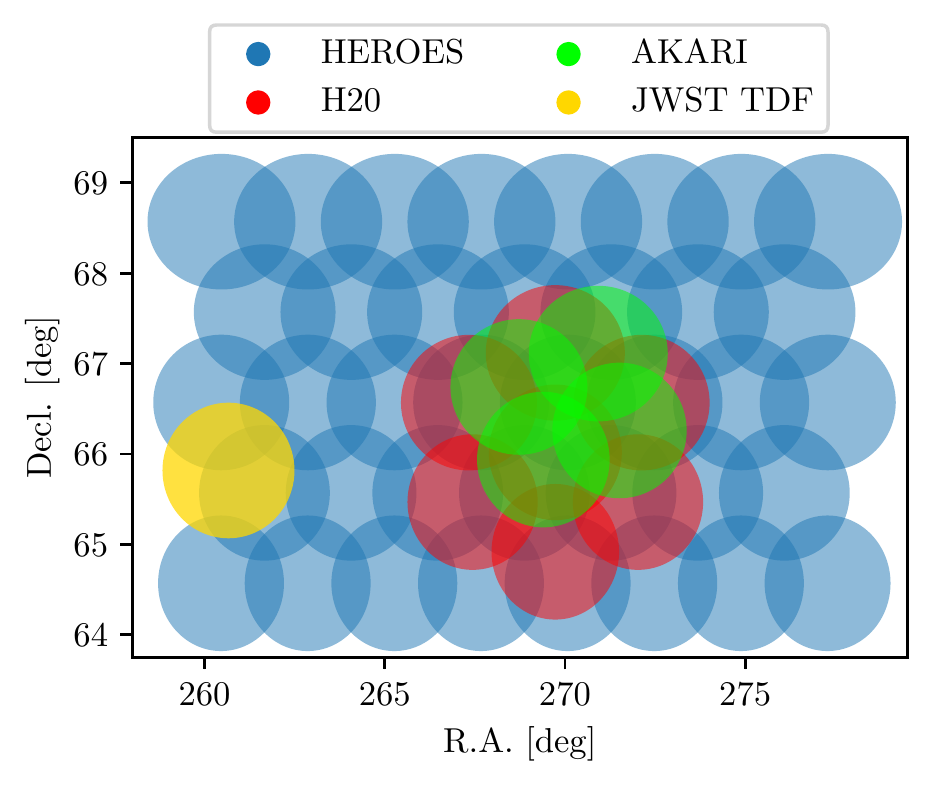}
\caption{Pointing centers and overlap for the HSC data in the NEP. Our HEROES pointings are shown 
in blue, the H20 pointings in red, and the AKARI pointings in green. Our pointing targeting the JWST TDF 
is shown in yellow. Each of the pointings used either the HEROES dither patten or something similar 
(see \S{\ref{sec:observations}}). Color shading shows pointing overlaps but does not explicitly indicate 
stacked imaging depth.}
\label{fig:pointings}
\end{figure}

We summarize the NEP HSC observations in Table~\ref{tab:obs} and illustrate the pointing centers in Figure~\ref{fig:pointings}. 
Our HEROES observations contributed 2574 targeted shots (a ``shot'' is a single HSC exposure consisting of 112 mosaicked CCDs).  We used a hexagonal grid of 38 pointings separated by 1\fdg0 to balance pointing overlap with the 1\fdg5 diameter HSC field of view (see Figure~\ref{fig:pointings}). In our pointings, we used a six point mosaic dither pattern with one central shot and five shots evenly distributed around a 2\farcm0 radius circle centered on the central shot \citep[see][their Figure~1]{songaila18}. In addition to our pointing grid, we conducted observations at a specialized pointing with the HSC-r2, HSC-z, and NB921 filters centered on the JWST TDF \citep{windhorst17,windhorst22}. Our HEROES observations were taken using the standard set of HSC broadband filters: HSC-g, HSC-r2, HSC-i2, HSC-z, HSC-Y, and the NB816 and NB921 narrowband filters (hereafter, referred to as $g$, $r$, $i$, $z$, $y$, NB816, and NB921).

In addition to the HEROES NEP observations, the AKARI-HSC survey \citep[][hereafter, AKARI]{oi21} previously observed 5.4~deg$^2$ of the NEP in $grizy$ broadbands to complement the infrared coverage of the AKARI satellite \citep{matsuhara06,murakami07,lee09}. The Hawaii Two-0 \citep[H20;][]{beck20} survey also combined parts of the HEROES raw data with new observations to provide $griz$ broadband coverage to complement the upcoming Euclid Deep Field North \citep{amendola13,amendola18}. We incorporated 181 archival shots in HSC $g$, $z$, and $y$ filters from the AKARI \citep{oi21} and 305 archival shots in $g$, $r$, $i$, and $z$ filters from H20 \citep{beck20} into our present HEROES data reduction. All together, these observations cover 44~deg$^2$ with full 7-band imaging.

\begin{deluxetable}{ccccccccc}[ht]
\setlength{\tabcolsep}{2pt}
\tablecaption{HSC Observations}
\label{tab:obs}
\tablehead{Filter\tablenotemark{a} & \multicolumn{2}{c}{HEROES} & \multicolumn{2}{c}{H20} & \multicolumn{2}{c}{AKARI} & \multicolumn{2}{c}{Total}\\ & Shots & Hours & Shots & Hours & Shots & Hours & Shots & Hours}
\startdata
HSC-g & 355 & 8.90 & 40 & 2.78 & 100 & 7.92 & 495 & 19.59\\
HSC-r2 & 208 & 9.02 & 60 & 3.33 & \nodata & \nodata & 268 & 12.35\\
HSC-i2 & 452 & 16.80 & 100 & 6.14 & \nodata & \nodata & 552 & 22.94\\ 
HSC-z & 735 & 24.21 & 105 & 7.06 & 29 & 1.85 & 869 & 33.11\\
HSC-Y & 241 & 10.78 & \nodata & \nodata & 52 & 4.55 & 293 & 12.23\\
NB816 & 235 & 11.33 & \nodata & \nodata & \nodata & \nodata & 235 & 11.33\\
NB921 & 348 & 20.86 & \nodata & \nodata & \nodata & \nodata & 348 & 20.86\\
\hline Total & 2574 & 101.79 & 305 & 19.31 & 181 & 14.32 & 3060 & 135.42\\
\enddata
\tablenotetext{a}{We refer to the HSC-g, HSC-r2, HSC-i2, HSC-z, and HSC-Y filters as $g$, $r$, $i$, $z$, and $y$ respectively throughout this article.}
\end{deluxetable}

\pagebreak
\section{Data Reduction}\label{sec:reduction}
We processed the HEROES dataset using the standard procedure provided by the \texttt{hscPipe} 
pipeline version 8.4 \citep{bosch18}. We performed this processing on the National Astronomical 
Observatory of Japan Large-scale Data Analysis System computing cluster with access provided 
from our September 2021 HSC observations. \texttt{hscPipe} is a comprehensive 
pipeline for the reduction and processing of HSC data that acts in 5 main steps: 
Bias/Dark/Flat/Fringe processing, Single visit processing, Moasicking, Coadding images, 
and Multiband analysis. In each of these steps, we used the default \texttt{hscpipe} parameters,
unless noted below. 

\begin{figure*}[ht]
\centering
\includegraphics[angle=0,width=\textwidth]{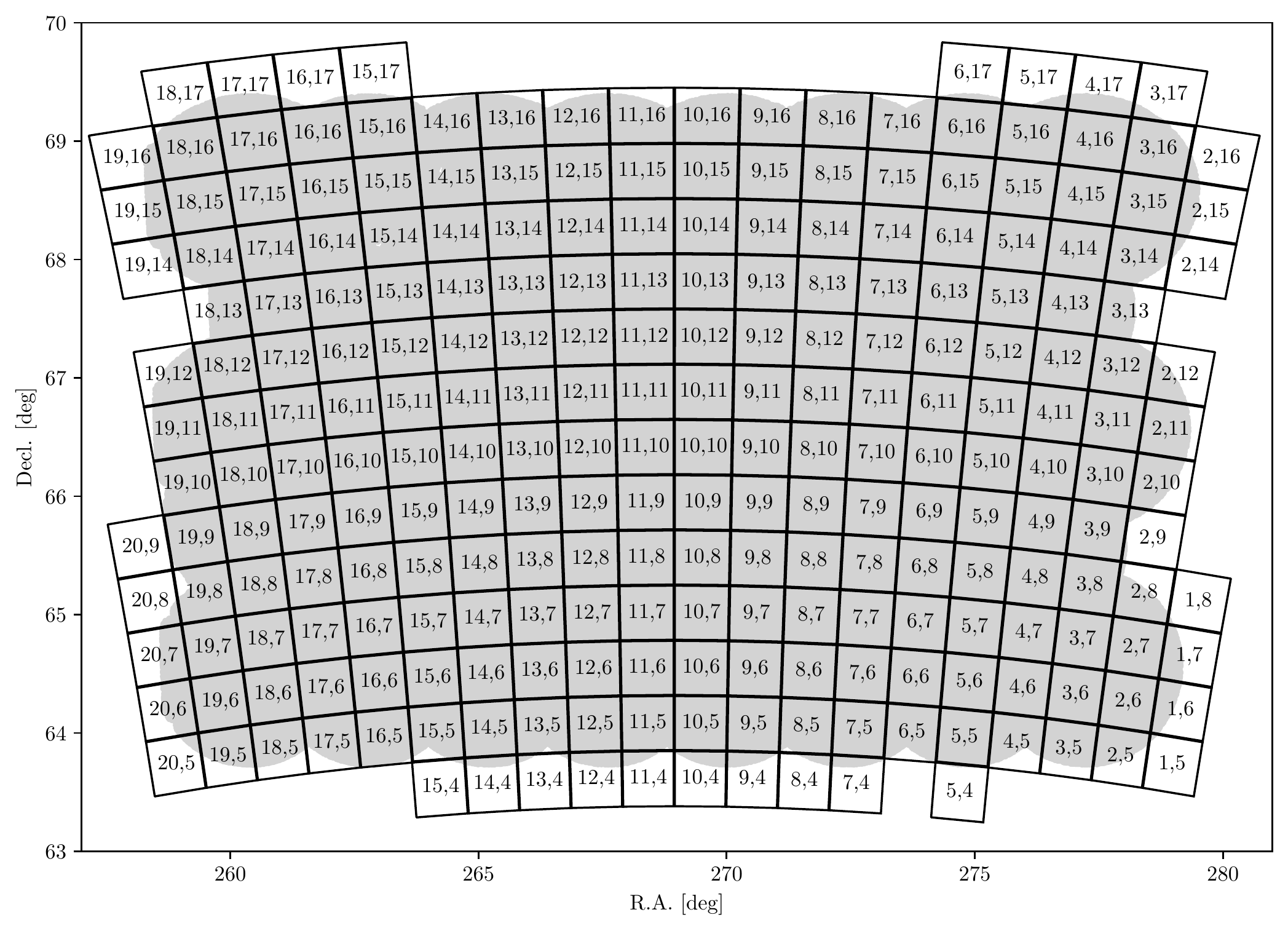}
\caption{Patch locations for the HSC data in the NEP. The bounding box of each patch is shown in black and 
labeled with its patch ID. The field imaging coverage is shown in the background in light gray shading.}
\label{fig:patches}
\end{figure*}

We assembled preprocessed Bias/Dark/Flat/Fringe data from the HSC calibration data archive\footnote{\url{https://www.naoj.org/Observing/Instruments/HSC/calib_data.html} - last accessed 2023 February 13} that matched all of the science data filters and observing runs. We performed single visit processing (detrending, WCS, and photometric calibration) on all individual CCD frames (103 science frames per shot) using the pipeline defaults. In the mosaicking step, we defined a custom tract and patch scheme. A \texttt{hscPipe} tract is an area of sky over which images are mosaicked and coadded with a common WCS solution. To produce data products of manageable size, a tract is divided into a grid of images called patches. We defined a single tract that encompassed the entire data coverage, and we divided it into a grid of $20\times14$ patches, each measuring 10,200 $\times$ 10,200 pixels to allow for overlap between adjacent patches. We show the position of the patches on the sky in Figure~\ref{fig:patches}. We used the \texttt{hscPipe} FGCM (Forward Global Calibration Method) to calibrate the data simultaneously in all seven filters to Pan-STARRS photometry and astrometry.

We executed the standard imaging coadding routines for each filter. We then ran the standard multiband analysis pipeline procedures to combine the detected source catalogs across the seven different filters for each patch and to perform forced photometry on the resulting combined catalogs. After completing this step, we noticed that the coadded images at the edges of the field included extrapolated background modeled pixels that extended beyond the science imaging coverage. These extrapolated pixels resulted in erroneous measurements of object fluxes during the multiband analysis step on our initial processing run. To remedy this, we masked all pixels outside of the science imaging coverage.
We re-ran the multiband measurements after this manual change and confirmed that our modification resolved the irregular photometric measurements.

\section{The Catalog}\label{sec:catalog}
We present the final HEROES catalog in three forms to best provide the maximum amount of available information, 
as well as convenient and manageable file sizes. 

First, we provide the forced photometry output catalogs from the multiband analysis step of \texttt{hscPipe}. These are organized by filter and patch using the naming format \texttt{forced\_src-\{filter\}-0-\{patch\}.fits}. The columns of these FITS tables are provided in the FITS headers and are summarized in the \texttt{hscPipe} documentation\footnote{\url{https://hsc.mtk.nao.ac.jp/pipedoc/pipedoc_8_e/tutorial_e/schema_multiband.html} - last accessed 2023 February 13}. These 1671 catalogs each contain 239 columns and total 211~GB (compressed). While these catalogs provide all of the available \texttt{hscPipe} forced photometry information, they are sometimes inconvenient for use in multifilter full field studies and contain many columns that are not useful for typical research applications. 

We also provide two catalogs that each include all of the objects in the dataset and select columns from the \texttt{forced\_src} catalogs. The first catalog (\texttt{HEROES\_Full\_Catalog.fits}, 25,445,387 objects, 172 columns, 12.4~GB) contains all of the columns described in the Appendix in Table~\ref{tab:columns}. The second catalog (\texttt{HEROES\_Small\_Catalog.fits}, 25,445,387 objects, 37 columns, 4.1~GB) is designed for more basic analyses or for machines with less system memory and only contains a subset of the full catalog's columns. It contains selected columns, as noted in the Appendix in Table~\ref{tab:columns}. 

In these catalogs, we converted \texttt{forced\_src} fluxes to AB magnitudes. To preserve the information for sources with negative measured fluxes, we report these values as negative magnitudes. For example; we report an object with a measured flux of $-3631\times10^{-11}$~Jy as an AB magnitude of -27.5. For objects that do not have imaging coverage in a given filter, or lack measured fluxes in a given filter, we report magnitudes of -99. 

These data products are all publicly available at Harvard Dataverse\footnote{\url{https://dataverse.harvard.edu/dataverse/heroes}}. 

\section{Data Quality}\label{sec:quality}

We tested the processed data quality by calculating 2\farcs0 diameter aperture magnitude depths across the field and in each filter with two different methods. In our first method, we placed 100,000 apertures randomly in each patch ($\sim$150 apertures per arcmin$^2$) and measured the flux in each aperture. We then divided the patch into 100 subregions and analyzed the apertures in each region separately. We used sigma-clipping to discard apertures that captured flux from field objects and took the standard deviation of the flux measurements from the remaining apertures. We converted this flux standard deviation to a magnitude to produce a 1$\sigma$ aperture magnitude depth, which we then converted to a 5$\sigma$ depth for the subregion by subtracting 1.75~mag ($-2.5\log_{10}5$). We report the overall median 2\farcs0 diameter aperture magnitude 5$\sigma$ depths as: $g$: 26.0, $r$: 25.7, $i$: 25.2, $z$: 24.7, $y$: 23.7, NB816: 24.3, NB921: 24.2. 

\begin{figure*}[ht]
\centering
\includegraphics[angle=0,width=\textwidth]{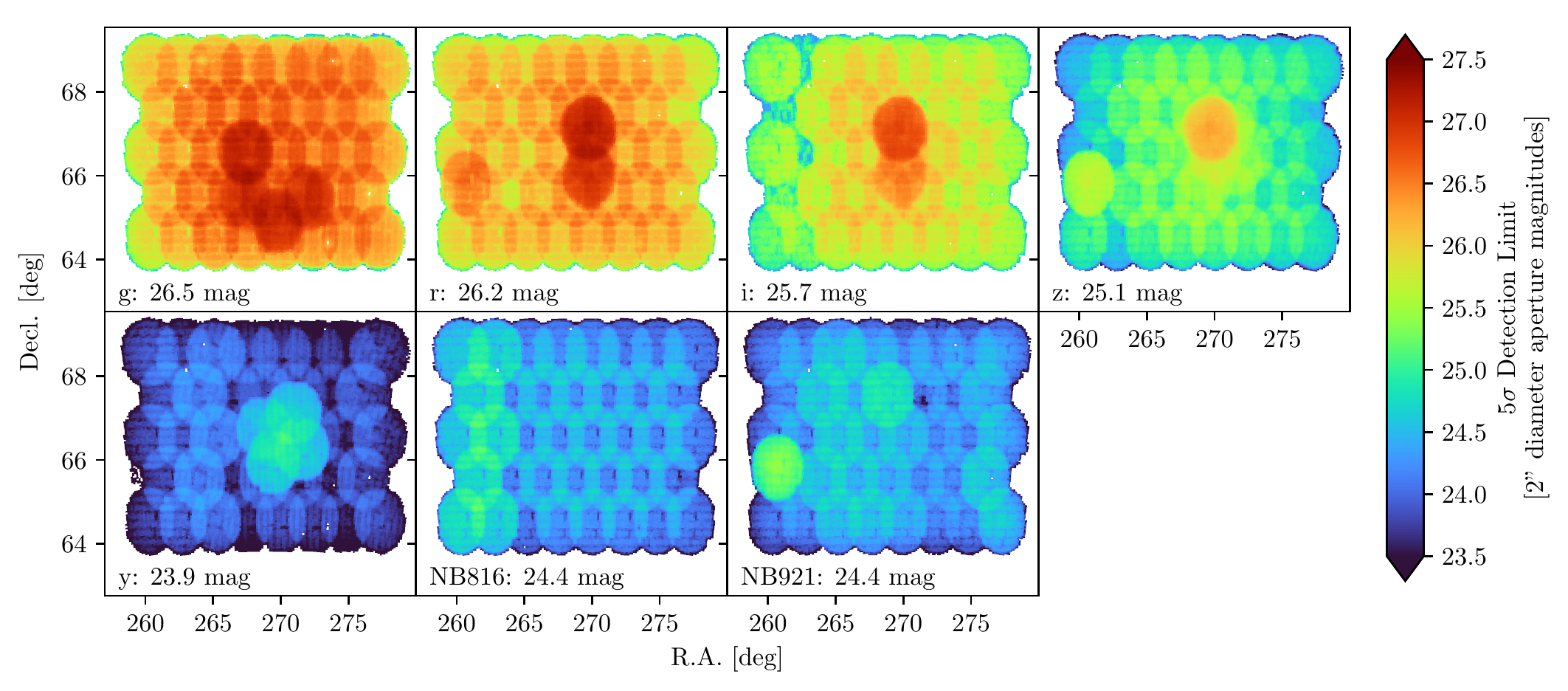}
\caption{5$\sigma$ detection limits for each filter across the survey field from the \cite{oi21} method. Note the depth increases in the centers of the $g,r,i,z,y$ imaging from the overlapping archival data and the significant depth increases in $r$, $z$, and NB921 near the JWST TDF (R.A.\ 17:22:47.896, Decl.\ +65:49:21.54) from our additional targeted pointing.
}
\label{fig:depth}
\end{figure*}

In our second calculation method, we adopted the methodology of \cite{oi21}. Here, we divided the survey into $150\times150$ subregions. In each subregion, we used the catalog provided fluxes and flux errors to select objects with a signal-to-noise of $\sim$5; ($5\pm0.1$). We then took the median magnitude of this S/N$\sim$5 population as the 5$\sigma$ depth for the subregion. For this method, we report the overall median 2\farcs0 diameter aperture magnitude 5$\sigma$ depths as: $g$: 26.5, $r$: 26.2, $i$: 25.7, $z$: 25.1, $y$: 23.9, NB816: 24.4, NB921: 24.4, and we show their variations across the field in Figure~\ref{fig:depth}. We attribute the 0.1--0.5 magnitude differences between the two methods to potential pattern noise effects or correlated variance that may not be fully sampled by the measurement routines in \texttt{hscPipe}, and/or to not fully cleaning contamination in the aperture method. The two methods may bracket the true limits.

\begin{figure*}[ht]
\centering
\includegraphics[angle=0,width=\textwidth]{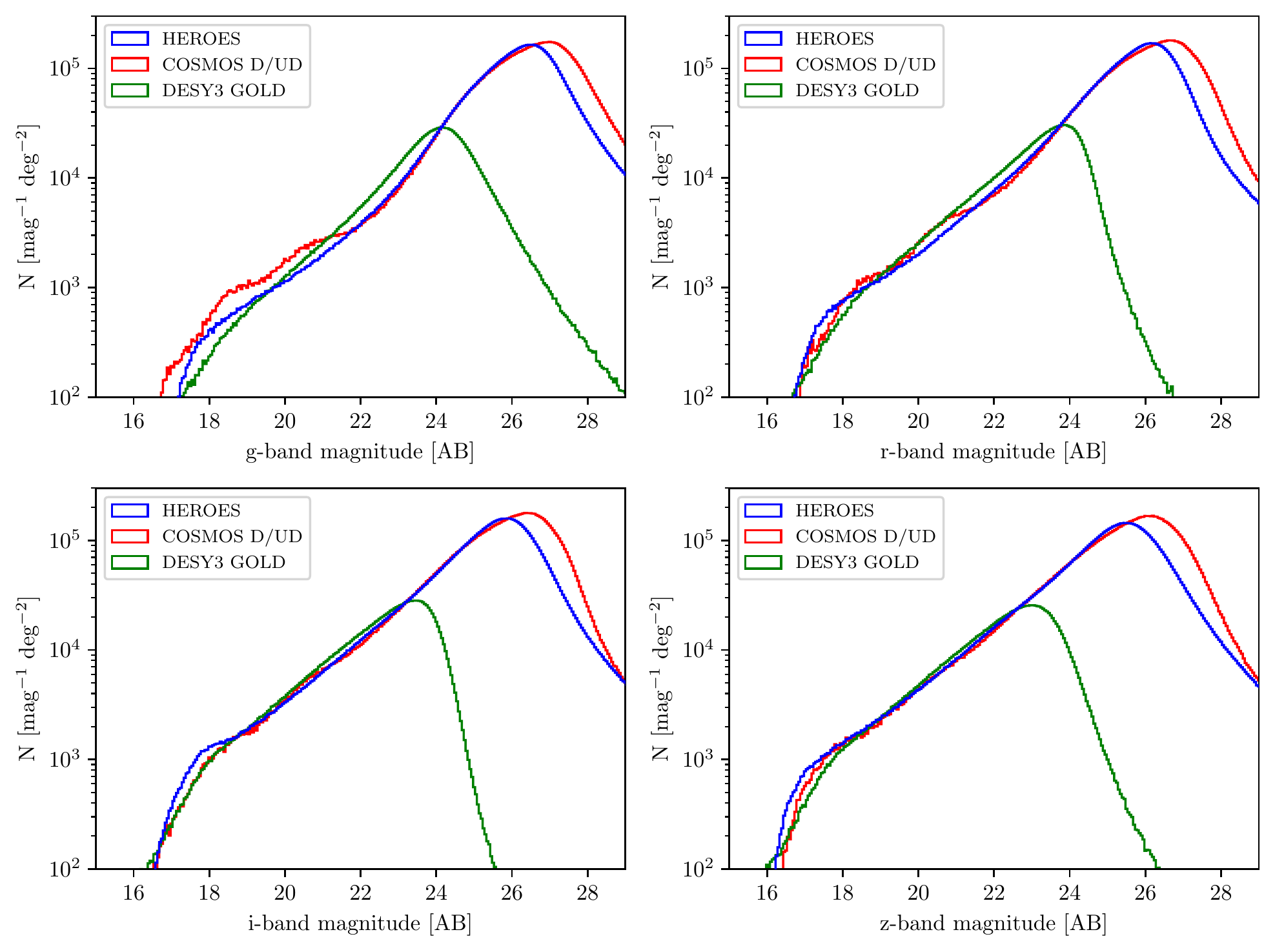}
\caption{Area density of sources in the $g$, $r$, $i$, and $z$ bands as a function of magnitude from HEROES (blue, 2'' diameter aperture magnitudes), COSMOS D/UD \citep[red, 2'' diameter aperture magnitudes,][]{aihara21}, and the DESY3 GOLD catalog \citep[green, single object fitted magnitudes,][]{sevilla-noarbe21,hartley22}. While the DESY3 catalog appears to be slightly overdense in the $g$ and $r$ bands, HEROES shows strong agreement with the area densities from COSMOS D/UD in all four bands and DESY3 in $i$ and $z$.}
\label{fig:density}
\end{figure*}

We tested the data quality by comparing the spatial number density of sources as a function of magnitude against other large surveys. To limit contaminating sources in our catalog for these comparisons, we applied a number of cuts to the catalog. First, we required all catalog objects (both stars and galaxies) to have the \texttt{is\_primary} flag. This flag selects objects that are not detected as blended composite objects. For each filter, we also removed all objects with the flag \texttt{\{filter\}\_base\_PixelFlags\_flag\_edge}. This cut has two main effects: First, it removes all sources that are outside of the science imaging coverage. Second, it removes bad sources near saturated pixels (e.g., near diffraction spikes from bright stars). After these cuts, we compared the area density of sources in our catalog to the HSC COSMOS Deep/UltraDeep (D/UD) catalog from HSC-SSP \citep[10.0 deg$^2$,][]{aihara21} and the Dark Energy Survey Year 3 GOLD catalog \citep[DESY3; 5347 deg$^2$;][]{sevilla-noarbe21,hartley22}. We use 2\farcs0 diameter aperture magnitudes from \texttt{hscPipe} for HEROES and COSMOS, and we use ``Single Object Fitted Corrected'' Magnitudes for DESY3 (2\farcs0 diameter aperture magnitudes are not provided in the DESY3 Gold Catalog). We show this comparison in Figure~\ref{fig:density}.

Across $g$, $r$, $i$, and $z$, HEROES (blue) shows excellent agreement with COSMOS D/UD (red). The DESY3 Gold catalog appears to be slightly overdense in the bluer $g$ and $r$ bands when compared to the other catalogs, but it shows good agreement in the $i$ and $z$ bands. Differences between the catalogs at the bright end are likely due to minor contamination by bright stars and image artifacts, while differences between the catalogs at the faint end are due to the different depths between the surveys. From these comparisons, we conclude that HEROES is consistent with other leading surveys that have well-calibrated photometry and low contamination. 

We further characterize the contamination rate in HEROES through visual inspections of 1000 sources drawn randomly from the filtered sample described above. In these inspections, we inspect $grizy$ thumbnails of the sources and look for diffraction spikes, glints, halos, or other visual artifacts that are detected as the source under inspection. Of the 1000 inspected sources, we find that only 27 (2.7$\pm$0.5\%) are impacted by visual artifacts. In most cases, the artifacts are the unsaturated tails of diffraction spikes or halo-like arcs from bright stars in the field. 

\section{Sample Selections} \label{sec:samples}
Here we demonstrate as examples two different sample selections using HEROES. These are the focus of our previous and upcoming research projects.  

\subsection{Narrowband Selection of $z\sim6$ LAEs}

We used a previous reduction of the HEROES data for our sample selections of $z\sim5$ and $z\sim6$ LAEs in \cite{taylor20,taylor21} and \cite{songaila22}. We now repeat these selections to verify the photometric consistency between the two reductions. 

\begin{figure}[ht]
\centering
\includegraphics[angle=0,width=\columnwidth]{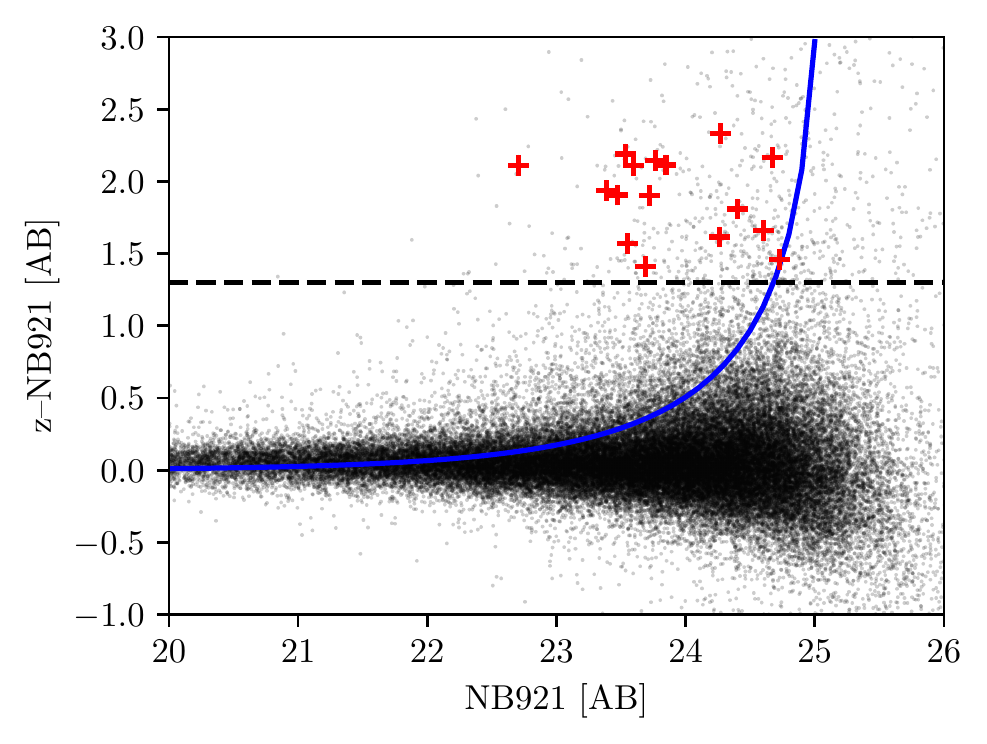}
\caption{Color excess vs.\ magnitude diagram for $z\sim6.6$ LAEs in HEROES. The black points are 1\% of the HEROES catalog that pass our initial source quality cuts. The blue curve is the narrowband excess significance cut \citep{sobral13}, and the dashed black line is the narrowband excess cut. The red crosses are the spectroscopically confirmed $z\sim6.6$ LAEs from \cite{taylor20} and \cite{songaila22}.}
\label{fig:NBselection}
\end{figure}

Our selection criteria are as follows. First, we remove sources with bad pixel, bright object, edge pixel, saturated center, cosmic ray, and interpolated center flags in the $i$, $z$, $y$, and NB921 filters. We then remove sources within 6\farcs0 of any \textit{Gaia} star with \textit{Gaia} $g$ magnitude less than 8. In Figure~\ref{fig:NBselection}, we show $1\%$ of the resulting clean sample as black points. 
From this clean sample, we next require strong detections (\texttt{n921\_detect = True}, \texttt{n921\_base\_SdssCentroid\_flag = False}, \texttt{n921\_base\_SdssShape\_flag = False}) in NB921 and 2$\sigma$ non-detections (\texttt{\{filter\}\_detect = False}) with forced aperture magnitude uncertainties greater than 0.5~mags in the $g$, $r$, and $i$ bands to enforce a strong Lyman break blueward of the Ly$\alpha$ line. We select a narrowband excess $z$ -- NB921 $>1.3$~mags. We also adopt the $\Sigma$ parameter from \cite{sobral13}. This parameter characterizes the significance of a narrowband excess above the uncertainties in the NB921 and $z$ source magnitudes and is given by
\begin{equation}
\Sigma = \frac{1-10^{-0.4(z-NB921)}}{10^{-0.4(27-NB921)}\sqrt{\sigma_{NB921}^2+\sigma_z^2}} \,,
\end{equation} 
where $z$ and NB921 are the AB magnitudes of $z$ and NB921, $\sigma_{\rm NB921}$ and $\sigma_z$ are the average $1\sigma$ image count rate uncertainties in 2\farcs0 diameter apertures in NB921 and $z$, and 27 is the magnitude zeropoint of the imaging data. In our source selection, we require $\Sigma>3$. We show the resulting cut in Figure~\ref{fig:NBselection} (blue curve).  

We then visually inspect cutouts of the remaining sources in stacked $gri$, $z$, $y$, and NB921 to reject sources with significant contamination from an elevated background, glints, diffraction spikes, transients, or other artifacts. This visual inspection is also helpful in rejecting sources that are not detected in $g$, $r$, or $i$ separately but are visually identifiable in stacked $gri$ cutouts. For a narrowband selection targeting objects with $z$ -- NB921 $>1.3$ and NB921$<24.25$, the above cuts produced a sample of 384 LAE candidates. After visual inspection, we reduced this sample to 63 candidates that showed no hint of emission in stacked and smoothed $gri$ cutouts, had compact morphologies, and were visually free of contamination. This significant reduction in candidates through visual inspection is primarily due to the ability of stacked $gri$ cutouts to detect low-redshift sources that do not show significant $gri$ emission in single filter observations. Furthermore, narrowband excess and Lyman break samples are more susceptible to objects with visual artifacts and contamination, as many forms of contamination may artificially emulate the narrowband excess criteria and non-detections in the bluer bands. Removing these sources through stricter magnitude and color cuts may risk reducing the sample completeness of the inherently rare bright $z>6$ LAEs, thus we use visual inspections to eliminate contaminating objects to ensure that our samples remain both complete and pure.

From these selection criteria (excluding the sources at NB921$<24.25$ that were not visually re-inspected), we completely recover the \cite{taylor20} and \cite{songaila22} samples (shown in Figure~\ref{fig:NBselection} as red crosses) and identify additional candidates for spectroscopic follow up (A.~Songaila~et~al.~2023, in prep).  These candidates and the recovered previous samples are uniformly distributed across the survey field with no obvious visual clustering or gradient beyond minor correlations with the imaging depth.

\subsection{Broadband Selection of Dropout Galaxies}

In order to test further the quality and science potential of the dataset, we also demonstrate a broadband dropout selection using the selection criteria and color-color cuts from \cite{ono18}. In their study, ``GOLDRUSH'', they selected $z\sim4,5,6,7$ galaxies using the dropout method with the UltraDeep, Deep, and Wide fields from HSC-SSP. These fields total 102.7~deg$^2$ in combined area, with the largest field (W-XMM) providing 28.5~deg$^2$ of coverage. We are currently working on a full comparison with the GOLDRUSH luminosity functions and clustering analysis \citep{harikane18}, and we summarize the preliminary galaxy selection results below.

\begin{figure*}[ht]
\centering
\includegraphics[angle=0,width=\textwidth]{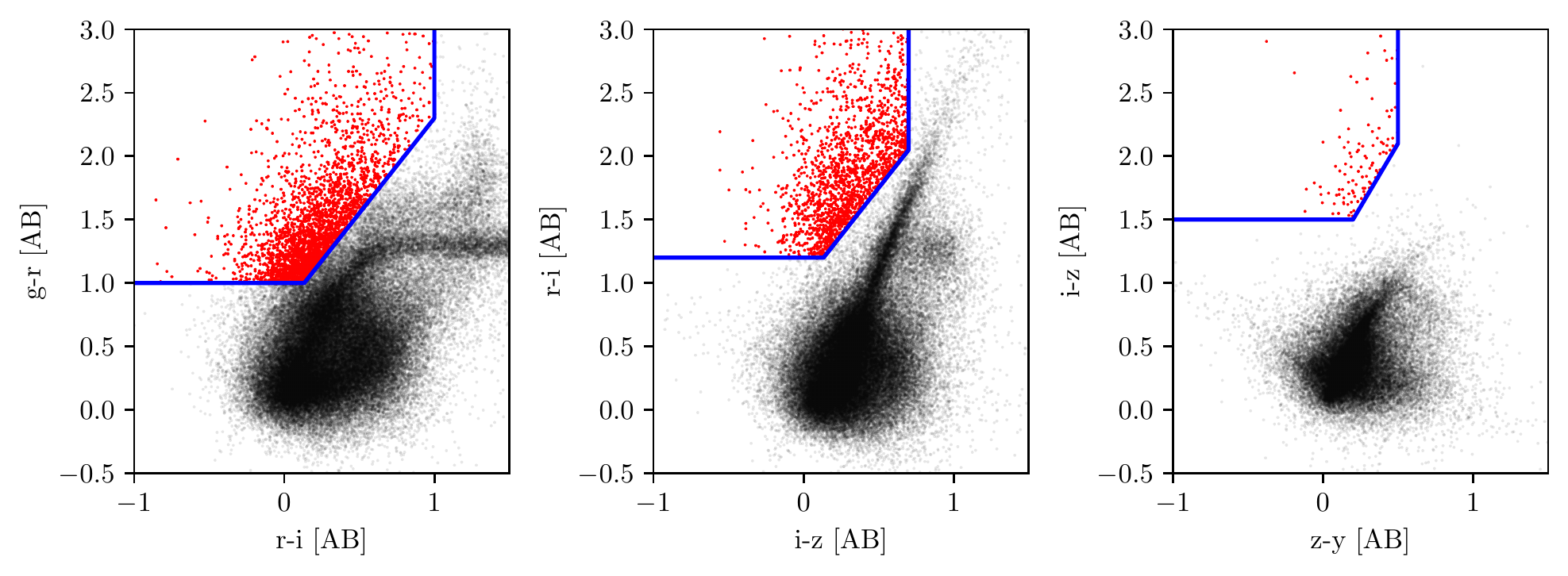}
\caption{Left: 1\% of the objects that passed the quality cuts but not the $g$-dropout color-color cuts (black points), and 1\% of the objects that also passed the $g$-dropout color-color cuts (red points). Center: 1\% of the objects that passed the quality cuts but not the $r$-dropout color-color cuts (black points), and 10\% of the objects that also passed the $r$-dropout color-color cuts (red points). Right: Center: 1\% of the objects that passed the quality cuts but not the $i$-dropout color-color cuts (black points), and 100\% of the objects that also passed the $i$-dropout color-color cuts (red points).}
\label{fig:dropout_selection}
\end{figure*}

As both the HEROES and HSC-SSP catalogs are produced by \texttt{hscpipe}, it is simple to adopt the \cite{ono18} selection criteria from their Table~2. In brief, they first required sources to have no bad pixel, bright object, edge pixel, saturated center, cosmic ray, or interpolated center flags in $grizy$. For each sample of $g$, $r$, $i$, and $z$-dropouts, they required non-detections in filters blueward of the dropout filter and strong detections in filters redward of the dropout filter. They then adopted color-color cuts from \cite{hildebrandt09} \citep[see][Equations 1--10]{ono18} to produce their initial dropout selections. We adopt the same cuts and show the color-color criteria and our results in Figure~\ref{fig:dropout_selection}.

In each panel of Figure~\ref{fig:dropout_selection}, we show a subset of the sources that pass the above described quality cuts as black points, and those that also pass the color-color dropout cuts as red points. For $g$, $r$, and $i$ dropouts, we find 295129, 18607, and 124 galaxies, respectively. This corresponds to $g$: 6700, $r$: 420, and $i$: 2.8 dropouts deg$^{-2}$. These surface densities are comparable to the densities from \cite{ono18} of $g$: 5300, $r$: 380, $i$: 5.2 deg$^{-2}$, and we attribute the offsets to differences in imaging depth and Poisson statistics. 

The surface densities of all three classes of dropouts are roughly uniform, with differences of no more than a factor of 2 over the HEROES field due primarily to differences in imaging depth both between bands and across the field.  We will refine these selections and compare our resulting luminosity functions and galaxy-galaxy angular correlation functions to the GOLDRUSH sample in A.~J.~Taylor~et~al.~2023, in prep.

\section{Summary}\label{sec:summary}
We present the complete photometric catalog from HEROES: a 44~deg$^2$ Subaru/HSC imaging survey of the NEP in $grizy$ broadbands and NB816+NB921 narrowbands. The catalog contains 25.4 million objects and is available in patch by patch, filter by filter \texttt{hscpipe} \texttt{forced\_src} format, as well as in two combined catalogs with selected columns. 

HEROES has enormous potential due to its overlap with other legacy, current, and future missions and surveys (e.g., AKARI, eROSITA, H20, S2CLS, NEPSC2, Spitzer, Euclid, JWST TDF). Outside of these complementary datasets, we are using HEROES to produce luminosity functions and angular correlation functions for $z\sim3-7$ dropout galaxies, as well as continuing to search for LAEs near the epoch of reionization.
We hope this public catalog release will enable new studies of galaxy evolution across cosmic time and provide complementary optical data for upcoming NEP surveys.

\acknowledgements
We thank the anonymous referee for their constructive report that helped us to improve this work.  

We gratefully acknowledge support for this research from NSF grants AST-1715145 (A.~J.~B) 
and AST-1716093 (E.~M.~H., A.~S.). We also gratefully acknowledge
the William F. Vilas Estate (A.~J.~T.) and a Kellett Mid-Career Award and a WARF Named 
Professorship from the University of Wisconsin-Madison Office of the Vice Chancellor for 
Research and Graduate Education with funding 
from the Wisconsin Alumni Research Foundation (A.~J.~B.).

This research is based on data collected at the Subaru Telescope, 
which is operated by the National Astronomical Observatory of Japan. 

Data analysis was carried out on the Multi-wavelength Data Analysis System operated by the 
Astronomy Data Center and the Large-scale Data Analysis System co-operated by the 
Astronomy Data Center and the Subaru Telescope.
We especially thank the HSC Software Help Desk for their rapid replies and helpful support 
during the data reduction process. 

We wish to recognize and acknowledge the very significant 
cultural role and reverence that the summit of Maunakea has always 
had within the indigenous Hawaiian community. We are most fortunate 
to have the opportunity to conduct observations from this mountain.

\facilities{Subaru Telescope}

\software{astropy: \cite{astropy:2013, astropy:2018}, hscPipe: \cite{bosch18}}

\bibliography{ref1}

\appendix
Here we show the combined catalog columns and descriptions in Table~\ref{tab:columns}.

\begin{deluxetable}{lcccl}[ht]
\tabletypesize{\scriptsize}
\setlength{\tabcolsep}{1pt}
\tablecaption{Catalog Columns}
\label{tab:columns}
\tablehead{\\Column & Unit & Type & Small Catalog & Description}
\startdata
id & \nodata & int64 & yes & \texttt{hscPipe} assigned unique ID number \\
ra & deg & float64 & yes & Right Ascension (J2000)\\
dec & deg & float64 & yes & Declination (J2000)\\
patch & \nodata & string & yes & Patch id\\
x & pixel & float32 & yes & x pixel location in mosaic\\
y & pixel & float32 & yes & y pixel location in mosaic\\
xx & pixels$^2$ & float32 & yes & xx pixel second moment\\
yy & pixels$^2$ & float32 & yes & yy pixel second moment\\
xy & pixels$^2$ & float32 & yes & xy pixel second moment\\
is\_primary & \nodata & bool & no & Object is not a blended object\\
num\_children & \nodata & float32 & no & Number of deblended child objects\\
\{$g|r|i|z|y|n816|n921$\}\_ap2(err) & mag & float32& yes & \{$g|r|i|z|y|n816|n921$\} 2\farcs diameter aperture magnitude (error)\\
\{$g|r|i|z|y|n816|n921$\}\_ap3(err) & mag & float32& no & \{$g|r|i|z|y|n816|n921$\} 3\farcs diameter aperture magnitude (error)\\
\{$g|r|i|z|y|n816|n921$\}\_ap4(err) & mag & float32& no & \{$g|r|i|z|y|n816|n921$\} 4\farcs diameter aperture magnitude (error)\\
\{$g|r|i|z|y|n816|n921$\}\_bkg(err) & mag & float32& no & \{$g|r|i|z|y|n816|n921$\} background magnitude (error)\\
\{$g|r|i|z|y|n816|n921$\}\_kron(err) & mag & float32& yes & \{$g|r|i|z|y|n816|n921$\} Kron magnitude (error)\\
\{$g|r|i|z|y|n816|n921$\}\_cmodel(err) & mag & float32& no & \{$g|r|i|z|y|n816|n921$\} CModel magnitude (error)\\
\{$g|r|i|z|y|n816|n921$\}\_blendedness & \nodata & float32& no & \{$g|r|i|z|y|n816|n921$\} \texttt{hscPipe} Blendedness factor\\
\{$g|r|i|z|y|n816|n921$\}\_detect & \nodata & bool & no & Detected in \{$g|r|i|z|y|n816|n921$\}\\
\{$g|r|i|z|y|n816|n921$\}\_base\_SdssCentroid\_flag & \nodata & bool & no & \{$g|r|i|z|y|n816|n921$\} centroid fitting error flag\\
\{$g|r|i|z|y|n816|n921$\}\_base\_SdssShape\_flag & \nodata & bool & no & \{$g|r|i|z|y|n816|n921$\} shape fitting error flag\\
\{$g|r|i|z|y|n816|n921$\}\_base\_PixelFlags\_flag\_edge & \nodata & bool & no & Object at edge of \{$g|r|i|z|y|n816|n921$\} imaging\\
\{$g|r|i|z|y|n816|n921$\}\_base\_PixelFlags\_flag\_bad & \nodata & bool & no & Bad pixel in \{$g|r|i|z|y|n816|n921$\}\\
\{$g|r|i|z|y|n816|n921$\}\_base\_PixelFlags\_flag\_interpolatedCenter & \nodata & bool & no & Interpolated pixel in \{$g|r|i|z|y|n816|n921$\}\\
\{$g|r|i|z|y|n816|n921$\}\_base\_PixelFlags\_flag\_saturatedCenter & \nodata & bool & no & Saturated pixel in \{$g|r|i|z|y|n816|n921$\}\\
\{$g|r|i|z|y|n816|n921$\}\_base\_PixelFlags\_flag\_crCenter & \nodata & bool & no & Cosmic ray pixel in \{$g|r|i|z|y|n816|n921$\}\\
\{$g|r|i|z|y|n816|n921$\}\_base\_PixelFlags\_flag\_bright\_object & \nodata & bool & no & Near bright object in \{$g|r|i|z|y|n816|n921$\}\\
\{$g|r|i|z|y|n816|n921$\}\_modelfit\_CModel\_flag & \nodata & bool & no & CModel fitting failed in \{$g|r|i|z|y|n816|n921$\}
\enddata
\end{deluxetable}

\end{document}